\documentclass[sigconf]{acmart}

\usepackage{booktabs} 
\usepackage{graphicx}
\usepackage{color}
\graphicspath{{../pdf/}{../jpeg/}}
\DeclareGraphicsExtensions{.pdf,.jpeg,.png}
\usepackage{subfigure}

\setcopyright{rightsretained}

\begin{document}


\copyrightyear{2019} 
\acmYear{2019} 
\acmConference[SSTD '19]{16th International Symposium on Spatial and Temporal Databases}{August 19--21, 2019}{Vienna, Austria}
\acmBooktitle{16th International Symposium on Spatial and Temporal Databases (SSTD '19), August 19--21, 2019, Vienna, Austria}\acmDOI{10.1145/3340964.3340986}
\acmISBN{978-1-4503-6280-1/19/08}

\title{Addict Free - A Smart and Connected Relapse Intervention Mobile App}

\author{Zhou Yang, Vinay Jayachandra Reddy, Rashmi Kesidi, Fang Jin }
\orcid{1234-5678-9012}
\affiliation{%
  \institution{Texas Tech University}
  \streetaddress{P.O. Box 1212}
  \city{Lubbock}
  \state{Texas}
}

\email{(zhou.yang, vinay.jayachandra, rashmi.kesidi, fang.jin)@ttu.edu}

\renewcommand{\shortauthors}{Zhou, Vinay, Rashmi, Jin}

\begin{abstract}
It is widely acknowledged that addiction relapse is highly associated with spatial-temporal factors such as some specific places or time periods. Current studies suggest that those factors can be utilized for better relapse interventions, however, there is no relapse prevention application that makes use of those factors. In this paper, we introduce a mobile app called ``Addict Free", which records user profiles, tracks relapse history and summarizes recovering statistics to help users better understand their recovering situations. Also, this app builds a relapse recovering community, which allows users to ask for advice and encouragement, and share relapse prevention experience. Moreover, machine learning algorithms that ingest spatial and temporal factors are utilized to predict relapse, based on which helpful addiction diversion activities are recommended by a recovering recommendation algorithm. By interacting with users, this app targets at providing smart suggestions that aim to stop relapse, especially for alcohol and tobacco addiction users. 

\end{abstract}

%
%


\maketitle


\section{Introduction}
Alcohol and tobacco are among the leading causes of preventable deaths in the United States. Approximately 46 million adults used both alcohol and tobacco in the past year. Alcohol and tobacco use may lead to major health risks when used alone and together. Due to an increase in the mortality rate of addicts, it is critical to help individuals recover from addiction. Though there are several treatments and rehabilitation methods available in the market, individual's opt to engage in self-management strategies to get rid of their addiction, whereas relapse is a top threat to self-management strategies. 
To prevent the problems in the relapse period and to stay away from addiction, one has to distinguish high-chance circumstances in which an individual is defenseless against relapse and to utilize psychological methods to stay away from addiction. Most of the previous studies on relapse prevention only concentrate on exploring efficient treatments, and presenting the consolidated amount of alcohol/tobacco they consumed, but lacking the dynamic monitoring, and individualized advice/recommendations, which is not much appealing to the users.

In this paper, we present a mobile app that closely monitors addicts' health statuses in their relapse period and assist them throughout the recovery span. Rather than motivating the users with inspirational quotes and treatments, this app enables them to engage themselves with a community. In our mobile app, the community comprises of people who have recovered completely, also who are in the process of recovery and addiction therapists. In particular, the following features make Addict Free mobile application, a smart and connected platform,  beneficial for various addicts.

\begin{itemize}
  \item Addict Free collects user's addictive behavioral data and generates recovery reports, and relapse prediction. 
  \item Addict Free intensively monitors users location using Geo-fence and provides personalized diversion recommendations. 
 
  \item Addict Free offers an interconnected support community, where users can share their experiences, post queries to overcome complications in the relapse period and offer suggestions for other users.
\end{itemize}

\begin{figure}[ht] 
    \centering
    \includegraphics[width=\linewidth]{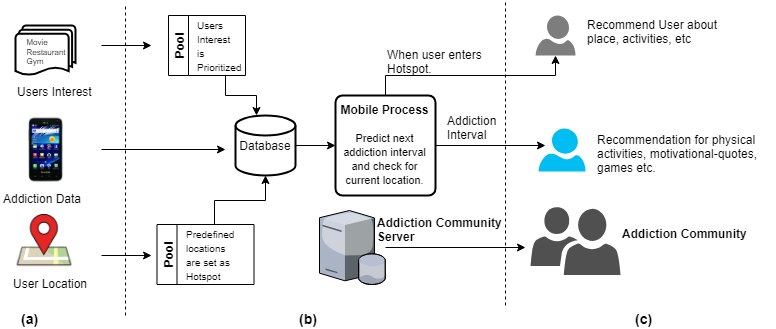}
    \caption{Framework of Addict Free mobile application}
    \label{fig:framework}
\end{figure}

\section{System framework}
 Figure~\ref{fig:framework} illustrates the system framework of Addict Free. The framework is divided into three main components: (a). A front-end component which collects data about spatial-temporal relapse pattern and diversion interest either manually or from social networking site; (b). A back-end component that includes not only multiple algorithms for predicting relapses that are associated with spatial factors and temporal factors, but also smart and personalized relapse diversions; (c). A relapse support community that enables users to share recovering experience and ask for suggestions. 
\section{General Modules}
\begin{figure}[!htbp]
	\centering
	\subfigure[]{
	    \includegraphics[width=0.3\linewidth]{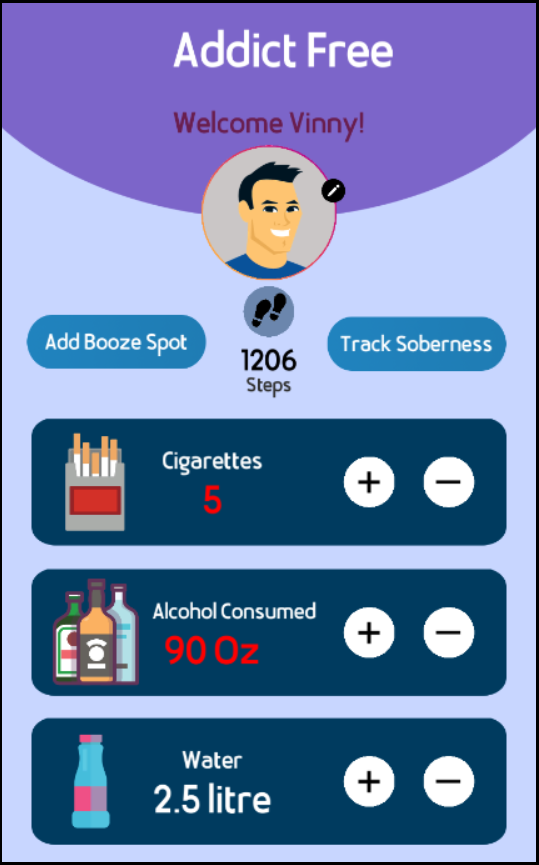}
	    \label{fig:home}
	}
    \subfigure[]{
		\includegraphics[width=0.3\linewidth]{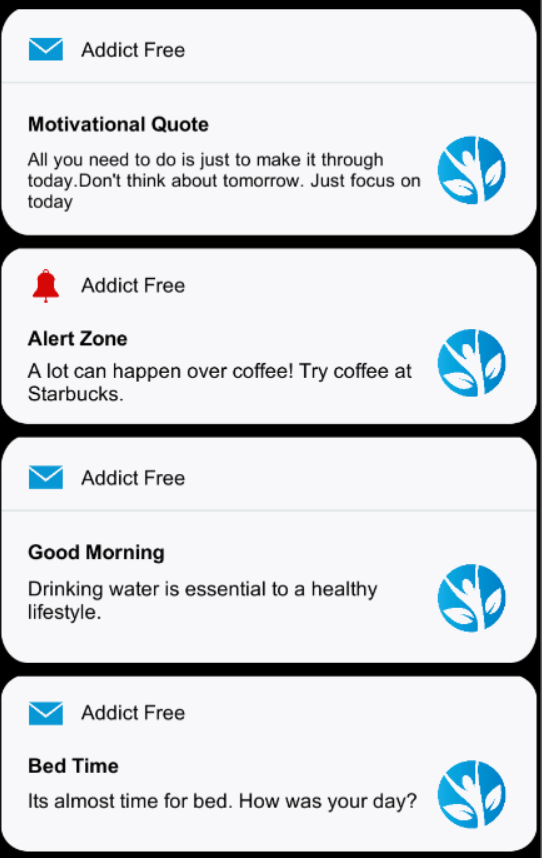}
		\label{fig:notification}
    }
    \subfigure[]{
		\includegraphics[width=0.3\linewidth]{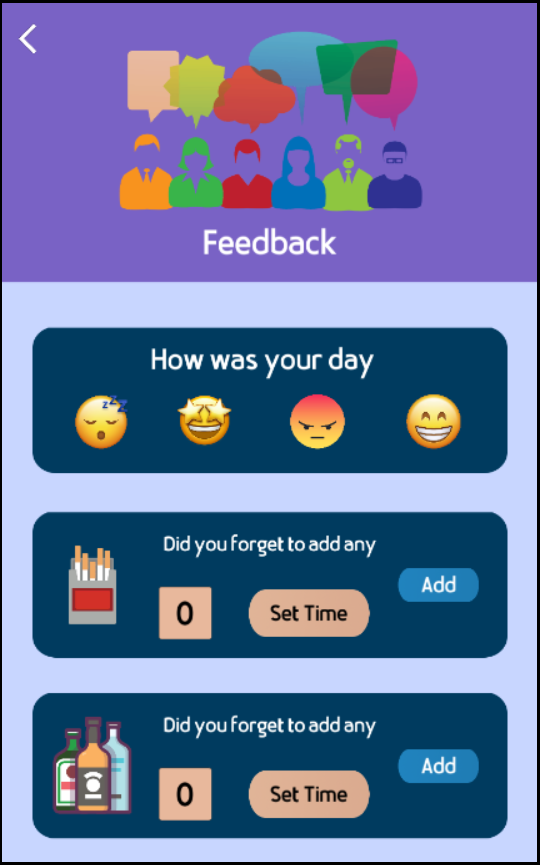}
		\label{fig:feedback}
    }
	\caption{(a). The mobile application interface where the user enters the data of alcohol consumed and cigarettes smoked, add their own alcoholic spot. (b). Different notifications to provide motivational quotes, recommendations in alert zones and to collect feedback. (c). Daily feedback screen implemented by a short survey.}
\label{fig:summary_notification_feedback}
\end{figure}


\paragraph{Dataset:}
The dataset used in this paper is collected from the application which is entered by each user as shown in Figure~\ref{fig:summary_notification_feedback}. The applications is still under development. The data type includes geo-spatial locations such as where users drink alcohol, time period at which user smokes, time at which user drinks, numeric numbers such as the quantity of cigarettes smoked and a number of ounces of alcohol user consumed. The collected data will be anonymized to remove personal information and stored in a database.

\subsection{Location Tracking}
\label{sec:location_tracking}

Geo-fence is created by storing latitude, longitude, and radius of harmful areas, which is used to track the user's present location\cite{yelne2015human}. Geo-fence empowers remote checking of geographic zones encompassed by a virtual fence (which is usually defined by the user) and automatic identifications when followed mobile devices enter or then again leave these zones. Geo-fence became important by supporting smart notifications in case the user enters or leaves a specific geographical area that is frequently associated with relapse
When a mobile device enters or leaves a particular location, a notification pops up to give a suggestion with a time constraint. For example, if an addict is spotted in any of the addiction-prone areas, the addict will be recommended with personalized diversion as shown in Figure~\ref{fig:notification}. 
 

\begin{figure*}[!htbp]
	\centering
    \subfigure[]{
		\includegraphics[width=0.3\linewidth]{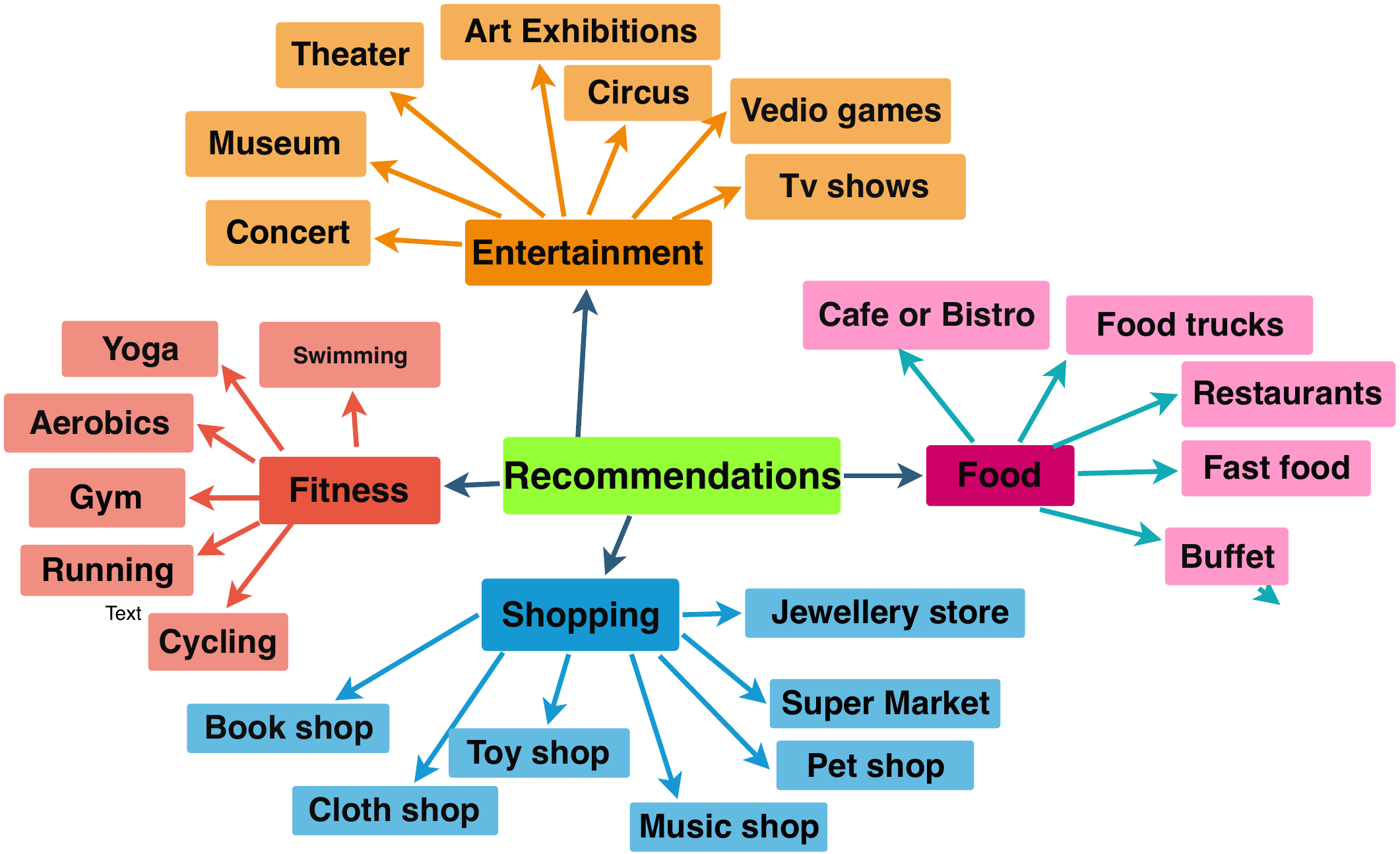}
		\label{word-cloud-irma}
	    \label{fig:personalized_intrest}
    }
    	\subfigure[]{
	    \includegraphics[width=0.2\linewidth]{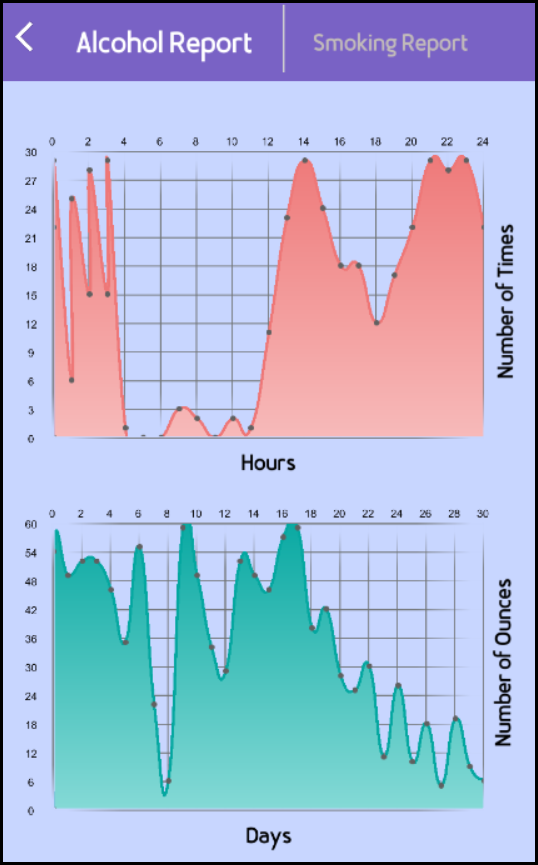}
	    \label{word-cloud-harvey}
	    \label{fig:alcohol_monthly}
	}
    \subfigure[]{
		\includegraphics[width=0.2\linewidth]{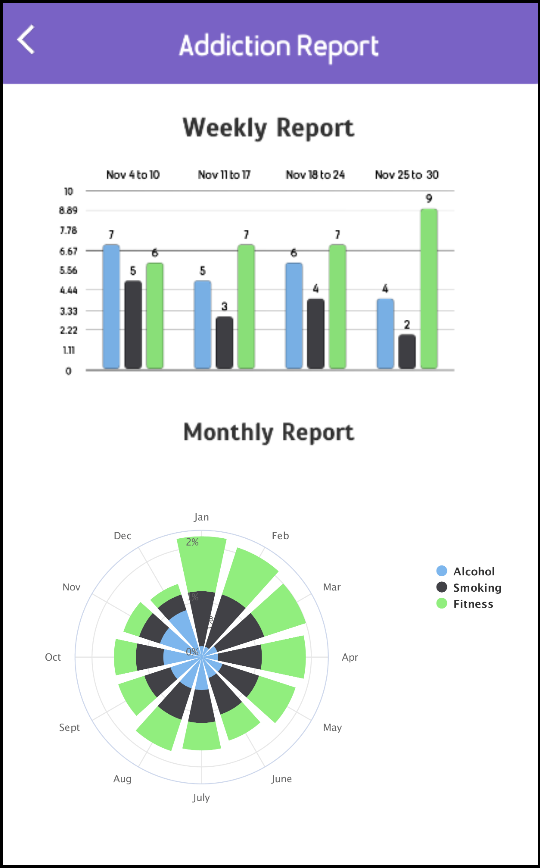}
		\label{word-cloud-sandy}
		\label{fig:Weekly}
    }
    \subfigure[]{
		\includegraphics[width=0.2\linewidth]{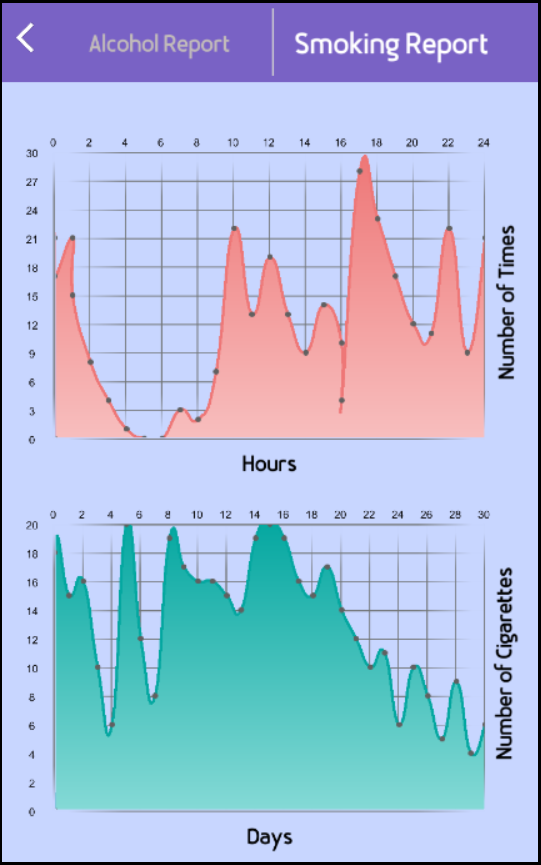}
		\label{word-cloud-irma}
		\label{fig:smoking_monthly}
    }
	\caption{(a). Personalized points of interest collected from the user to provide recommendations; (b).Snapshot of user's statistics about the number of times drank in a day and Number of ounces of alcohol consumed in a month; (c). Weekly statistics of alcohol, smoking and fitness and progress towards health; (d).Snapshot of user's statistics about the number of times smoked in a day and Number of cigarettes consumed in a month. }
\label{fig:summary_notification_feedback}
\end{figure*}

\begin{figure*}[!htbp]
	\centering
	\subfigure[]{
	    \includegraphics[width=0.2\linewidth]{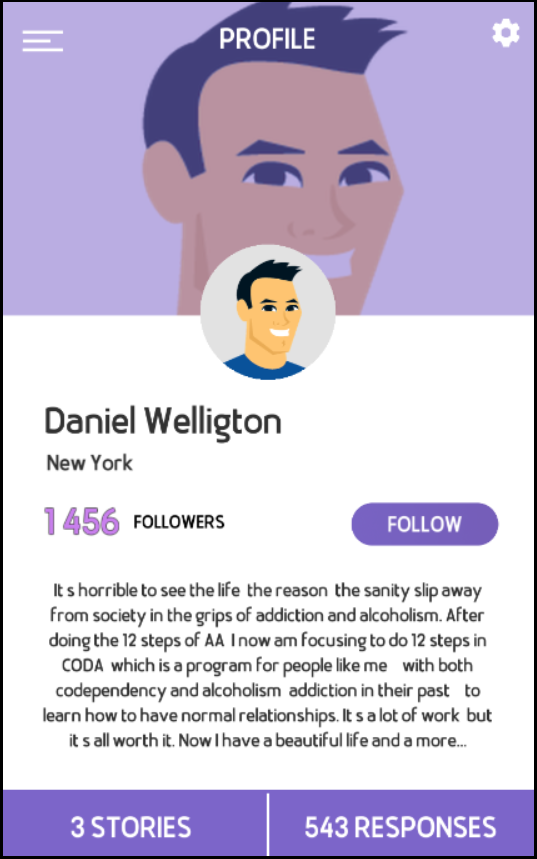}
	    \label{word-cloud-harvey}
	    \label{fig:profile}
	}
    \subfigure[]{
		\includegraphics[width=0.2\linewidth]{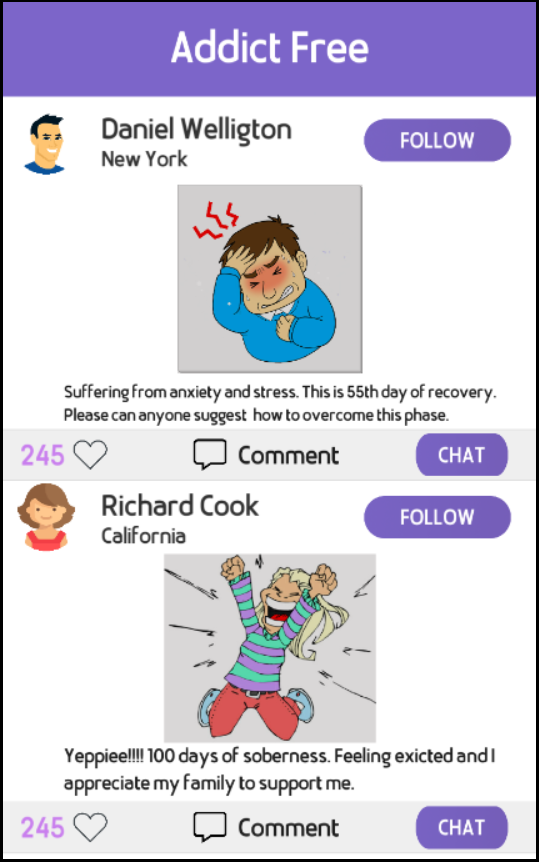}
		\label{word-cloud-sandy}
		\label{fig:blogpost}
    }
    \subfigure[]{
		\includegraphics[width=0.2\linewidth]{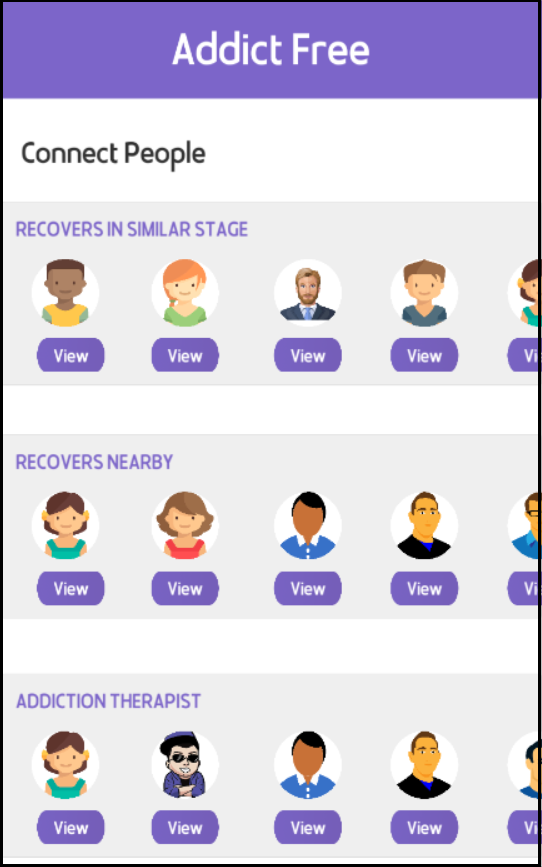}
		\label{word-cloud-irma}
		\label{fig:connections}
    }
     \subfigure[]{
		\includegraphics[width=0.2\linewidth]{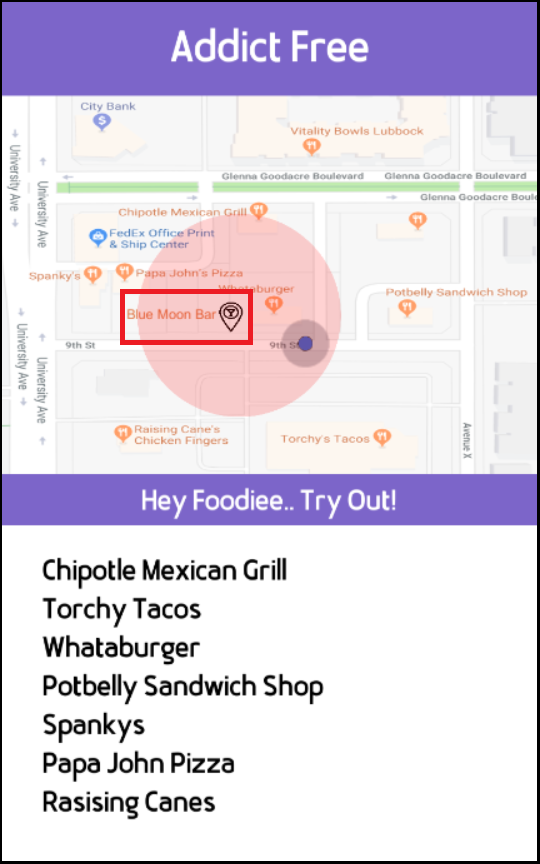}
		\label{word-cloud-irma}
		\label{fig:recommendation}
    }
	\caption{(a). Addict's profile screen; (b). A blog post made by a user which is visible to other users in the community; (c). Addiction community connectivity options provided to the user based on vicinity and stage; (d). Snapshot of recommendations provided to a user(whose point of interest is food) when the user enters into an alcoholic spot.}
\label{fig:community_social}
\end{figure*}

\subsection{User Notification}
\subsubsection{Alert zone for alcohol consumption}

Highlights of various places would directly or indirectly influence individuals life and work\cite{tunstall2004places}. These places have an influence on an individual`s addiction or behavior. There is a need to identify those likely places for a particular individual to divert or make the individual out of focus for their addiction. This can be achieved by providing a diversion in those areas. A notification pops up when a user enters public alcoholic locations or user specific alcoholic spots as shown in Figure~\ref{fig:notification} . Based on the interests of the user, a place nearby to the spot is suggested along with the alert.


\subsubsection{Diverting notifications} 
The idea behind diversion techniques is simple: if users focus on something (i.e, cravings for tobacco), their needs will seem more intense. If they were distracted, they can trick their mind into `forgetting' the craving and it will pass. This behavioral observation can be utilized for stopping smoking and drinking because cravings rarely last for longer than a couple of minutes. Specifically, the craving is diverted by a pop-up notification that is personalized according to the user profile and a couple of alternative activities that are useful for relapse diverting are also incorporated into the notification. As shown in Figure~\ref{fig:notification}, notifications will pop up 10 minutes before the time when the relapse is most likely to occur.

\subsection{Recovering Summaries}
Soberness and relapse states are also tracked on different time granules, which enables app users to have a better understanding of their recovering statuses.
A Weekly recovering summary is presented in Figure~\ref{fig:Weekly}. It depicts consolidated statistics of an individual's urge towards recovering on a weekly basis with a normalized score of range 1 to 10. For each day, three scores are built for alcohol, smoking and fitness respectively. Plots in Figure~\ref{fig:alcohol_monthly} and ~\ref{fig:smoking_monthly} depict the user's trend of consumption of alcohol/tobacco over a month. One plot depicts a number of times the user consumed alcohol/tobacco at an average time, and the other depicts the number of ounces of alcohol/number of cigarettes an individual consumed in a day for about a month.

\paragraph{User Recommendations and Feedback:}
User interests are collected initially to provide personalized recommendations as shown in Figure~\ref{fig:personalized_intrest}. Interests are subcategorized from the main themes which help our application to provide more personalized recommendations according to their interests and provide adequate ways to divert the user from his addiction and the places in which he is more prone to addiction.Every individual thinks of going back to a particular time in a day to complete things which they have left around. The user might not enter the data about alcohol or cigarette he consumed. Stress levels also have an impact on the day of the user. Considering all these, feedback is collected at the end of the day. Screenshot of feedback is shown in Figure~\ref{fig:feedback}.


\subsection{Addiction Community}

Addiction recovery calls for not only medical treatment such as replacement therapy but also support from families, medical professionals and communities. The function of community support is also incorporated into the Addict Free application. This community support function groups users and therapist with a similar background. Thus, it enables people to take suggestions from the ones who are a similar status and background. As shown in the  Figure~\ref{fig:profile}, Figure~\ref{fig:blogpost} and Figure~\ref{fig:connections}, recovering users and therapists with the similar background are grouped to enable potential connections between them. Also, users are allowed to post questions and other users in the community can provide suggestions by commenting or chatting directly with the post authors.

\section{Methods}
\subsection{Relapse Prediction}
\begin{figure}
    \centering
    \includegraphics[width=0.8\linewidth]{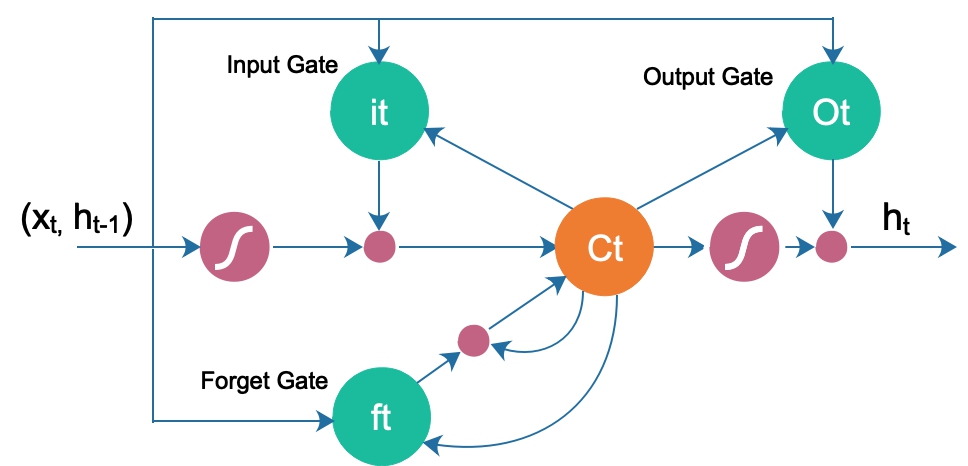}
    \caption{LSTM Architecture.}
    \label{fig:lstm_architecture}
\end{figure}

Addict Free incorporated algorithms for relapse prediction that is fundamental for relapse diversion and intervention~\cite{yang2019relapse}. Given a group of users $A=\{a_1, a_2,a_3,\dots\}$, and their corresponding feature variable $F_j=<X_1,\dots,X_m>$ and relapse label  $Y_j \in [0,1]$, a prediction algorithm is implemented to learn a function $f:R^m \to R^1$, such that the prediction errors are minimized.

\begin{equation}
    \label{equ:loss}
   args= \arg \mathop {\min }\limits_\Psi  \sum\limits_{j = 1}^m {||f({X_1,\dots,X_m}) - {Y_j}{||_2}}
\end{equation}
where $X_j$ and $Y_j$ are feature variable and relabel label respectively. The classifier is implemented in a LSTM Model  with time series data. To capture the real-time dynamics of input, Addict Free utilizes a LSTM model for relapse time series prediction due to its well-handling ability of long and short term time dependency. The model is also utilized in other types of prediction as in \cite{nguyen2018smart} and \cite{liang2018multi, nguyen2018spatial, nguyen2018forecasting}. Figure~\ref{fig:lstm_architecture} shows the basic structure of LSTM. It has an input gate $i_t$, output gate $o_t$, forget gate $f_t$ and memory cell $C_t$. Equations ~\ref{equ:lstm} shows how to update the output values each step. 

\begin{equation}
    \label{equ:lstm}
    \begin{aligned}
       f_t = {g(W_f}{.x_t+U_f.h_{t-1}+b_f)},\\
       i_t = {g(W_i}{.x_i+U_i.h_{t-1}+b_i)},\\
       c_t = {f_t.c_{t-1}+i_t.k_t},\\
       o_t = {g(W_o.x_t+U_o.h_{t-1}+b_o)},\\
       h_t = {o_t.tanh(c_t)}.
    \end{aligned}
\end{equation}
Where $x_t$ is the input vector and $g$ is the activation function such as ReLU or Sigmoid function. $W$ is the weight vector. 

The proposed model ingests multiple variables, including smoking (drinking) time, the number of cigarettes smoked (the amount of alcohol consumed). The model can extract the hidden patterns from these variables and output the probability of relapse in the next hour. For accuracy, previous 30-day data is utilized to predict the probability of relapse. Once the relapse probability for each hour is available, Addict Free will notify users with some diverting activities that may trick users' mind and lead to a successful relapse intervention.

\subsection{Monitoring and Notifying User}
Since relapse is highly associated with locations such as bars, it's also crucial to divert app users when they are at such locations where they used to drink (smoke). Addict free also provides notifications to divert users form location-based relapse. As mentioned in Section~\ref{sec:location_tracking}, Addict Free uses Geo-fence to identify and provides diverting notifications when a user enters or leaves such Geo-fences with a time and radius constraint. A Geo-fence state and transition can be annotated with duration constraints. These constraints specify that a mobile device needs to remain within a Geo-fence or remain in motion between Geo-fences for a limited duration out of a duration interval $[l_{min},l_{max}]$, respectively $(l_{min}, l_{max}) \in D$. $l_min$ defines the minimum duration and $l_max$ the maximum duration allowed \cite{rodriguez2014geofencing}. This allows our application to accurately confirm the existence of a user in a particular Geo-fence before leaving the location or entering into another Geo-fence. Thereby, the following conditions need to be fulfilled for Geo-fence-related duration constraints
\[
\forall (l_min,l_max) \in D | l_min \in \mathbb{R}^{\geq 0} \wedge lmax \in \mathbb{R}^{ > 0} \wedge l_min < l_max
\]
for duration constraints related to transitions
\[
\forall (l_min,l_max) \in D | l_min \in \mathbb{R}^{\geq 0} \wedge l_max \in \mathbb{R}^{> 0} \wedge l_min \leq l_max
\]

If a Geo-fence state is not annotated with a duration constraint, the allowed duration is implicitly assumed to be arbitrary, $d \in R^{>0}$ . For a transition without duration constraints, the allowed duration can be arbitrary but including zero as well, $d \in R^{\geq0}$ . In case a transition is annotated with $(0, 0) \in D$, a mobile device can e.g. enter $g_2$ immediately after leaving $g_1$ where $g_1$ and $g_2$ are two Geo-fence areas.

Recommendations are provided to users based on their interests like food, fitness, shopping, entertainment, etc., collected from their profile information. Figure~\ref{fig:recommendation} shows how personalized notification works. For example, an Addict Free user with thee drinking problem, entered into a Geo-fence area such as bar, then the user is notified to try out nearby food places which the user would be interested in according to Addict Free database. 

\section{Summary}
We developed a mobile application and relapse intervention assistant to help users stay clean from alcohol and smoking addiction. The platform incorporates a variety of data to provide insight into trends of smoking and alcohol consumption on a daily, weekly and monthly basis. Those trends allow users to comprehend their activities throughout the span of recovery. Also, spatial and temporal factors are utilized to predict the most likely relapse locations and time period. Embracing the Geo-fence technique for monitoring users' locations makes Addict Free progressively effectively prevent location-based relapse. Addict Free has a unique way of recommending diversions, which also consider the addict's personal preferences. Besides, this app builds a smart and connected addiction-free community where users share relapse prevention experience. Instead of struggling with addiction individually, users here can benefit from community support that is working towards the common goal to stay clean. Addict Free additionally encourages addicts to share involvement and get proposals from addiction therapists, which helps them to overcome complexities encountered during the relapse period.

\section{Acknowledgement}
This work was supported by the U.S. National Science Foundation under Grant CNS-1737634.

\vspace{-3mm}
\bibliographystyle{ACM-Reference-Format}
\bibliography{references}

\end{document}